\documentclass[fleqn,12pt,twoside]{article}
\usepackage{espcrc1}

\usepackage{graphicx}
\usepackage[figuresright]{rotating}

\newcommand{\AmS}{{\protect\the\textfont2
  A\kern-.1667em\lower.5ex\hbox{M}\kern-.125emS}}

\title{Properties of few-body systems in relativistic quantum mechanics and 
constraints from transformations under Poincar\'e space-time translations}

\author{B. Desplanques\address{Laboratoire de Physique Subatomique 
et de Cosmologie \\(UMR CNRS/IN2P3-UJF-INPG),  
F-38026 Grenoble Cedex, France}\thanks{desplanq@lpsc.in2p3.fr}   }

\begin{document}

\maketitle

\begin{abstract}
Different approaches have been applied to the calculation of form factors 
of various hadronic systems within relativistic quantum mechanics. In a one-body 
current approximation, they can lead to results evidencing large discrepancies. 
Looking for an explanation of this spreading, it is shown that, 
for the largest part, these discrepancies can be related 
to a violation of Poincar\'e space-time translation invariance. 
Beyond energy-momentum conservation, which is generally assumed,  
fulfilling this symmetry implies specific relations that are generally ignored. Their
relevance within the present context is discussed in detail both to explain 
the differences between predictions and to remove them.
\end{abstract}

\section{INTRODUCTION AND MOTIVATIONS FROM FORM FACTORS}
The implementation of relativity within quantum mechanics has been 
the object of extensive studies these last years, especially with regard 
to the prediction of properties of various hadronic systems, such as form factors. 
These ones could be a test for the underlying dynamics but this program 
supposes that accounting for relativity is under countrol. Looking at
predictions, which are generally based on a single-particle current, 
it is found that they can considerably differ, depending on the form
\cite{Dirac:1949cp} and the kinematics used for their calculation. 
The effect is especially large when the mass of the system is small 
in comparison of the sum of the constituent ones (pion for instance). 
The study of a theoretical model with a simple dynamics can give insight 
on the relevance of a given approach. It does not necessarily provide 
an argument why the results are so spread in some cases. It is the aim 
of the present contribution to bring the attention on the role, 
with this respect, of Poincar\'e space-time translation invariance. 
This symmetry implies energy-momentum conservation, which is generally 
assumed, but, within relativistic quantum mechanics (RQM), it also 
supposes some constraints. These ones are rarely considered 
and their fulfillment requires that the contribution of many-body currents 
be included.

In the second section, we remind results obtained in different approaches 
for the charge form factor of the ground state in a simple theoretical 
system. The third section is devoted to the role of Poincar\'e space-time 
translation invariance. The violation of the expected constraints is related 
to the discrepancies between various approaches. Results corrected 
for this violation are presented. Some prospects are given 
in the fourth section. 
\section{FORM FACTORS FROM A SIMPLE THEORETICAL MODEL}
We here remind results \cite{Desplanques:2004sp} obtained for the charge 
form factor of the ground state in the Wick-Cutkosky model. 
This one consists of scalar particles exchanging a massless scalar meson. 
It offers the advantage to be solvable and the resulting
Bethe-Salpeter amplitude can easily be used to calculate form factors. 
This represents our ``experiment". The corresponding form factors 
can be calculated in RQM approaches using the solution of a mass operator. 
This one is obtained from a fit to the lower-states 
masses, while fulfilling minimal requirements \cite{Amghar:2002jx}. 
The total and constituent masses are chosen as for a pion-like system. 
The detailed expression of form factors, which can be found in Ref. 
\cite{Desplanques:2004sp}, is common to all cases. It is 
in particular noticed that the current has a one-body form 
and  assumes the most natural choice, $<|J^{\mu}|> \propto (p_i+p_f)^{\mu} $
(with $p^0=+\sqrt{m^2+p^2}$).

\begin{figure}[htb]
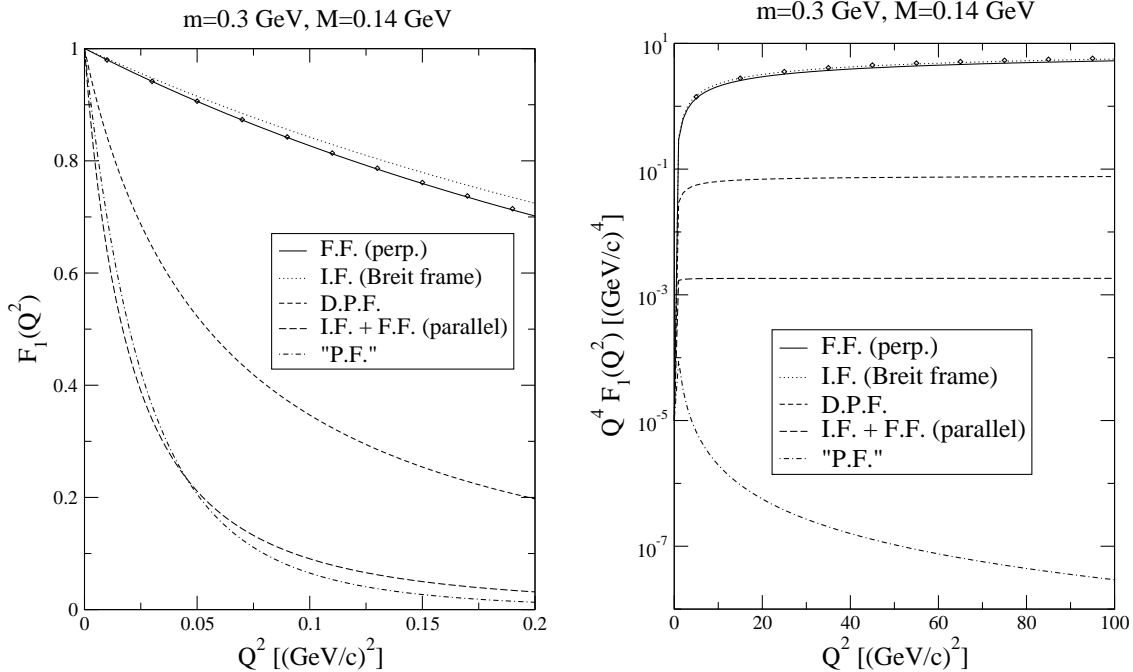

\includegraphics[width=0.45\textwidth]{DesplanquesB-fig1a.eps}
 \hspace*{4mm}\includegraphics[width=0.45\textwidth]{DesplanquesB-fig1b.eps}
\caption{Charge form factor at low and high $Q^2$ in different forms 
of relativistic quantum mechanics (see text for details). The ``experiment"
is represented by small diamonds.\label{fig1}}
\end{figure}

Numerical results are presented 
in Fig. \ref{fig1}, at small $Q^2$ to show the sensitivity 
to the charge radius and at high  $Q^2$ 
(multiplied by $Q^4$) to evidence the asymptotic behavior. 
The approaches under consideration involve:\\
- the front form with $q^{+}=0$ (``perpendicular" kinematics: 
F.F.(perp.)), \\  
- the instant form in the Breit frame (``perpendicular" kinematics: 
I.F.(Breit frame)), \\  
- the ``point-form" (Bakamjian \cite{Bakamjian:1961}, 
Sokolov \cite{Sokolov:1985jv}: ``P.F."), \\  
- the point-form (Dirac inspired \cite{Desplanques:2004rd}: D.P.F.), \\  
- the front-form (``parallel" kinematics, 
$(\vec{P}_i\!+\!\vec{P}_f) \parallel \vec{Q} \parallel \vec{n}$: F.F.(parallel)), \\  
- the instant-form (``parallel" kinematics, 
$(\vec{P}_i\!+\!\vec{P}_f) \parallel \vec{Q}$, 
$|\vec{P}_i\!+\!\vec{P}_f| \rightarrow \infty$:
 I.F.(parallel)). 

In comparison to the  ``experiment", it is noticed that the standard front 
and instant forms  (F.F.(perp.) and I.F.(Breit frame)) do well. 
In the other cases, a large sensitivity to the smallness of the total mass, 
$M$, is observed, the charge radius scaling like $M^{-1}$. 
It is also noticed that fulfilling  Lorentz invariance (``P.F." and D.P.F.) 
does not ensure good results. Altogether, these observations raise 
a double question: 
What is the reason for large discrepancies? 
Why some approaches do better than other ones?

\section{CONSTRAINTS FROM POINCAR\'E SPACE-TIME TRANSLATION INVARIANCE}
Poincar\'e space-time translation invariance is a symmetry whose 
consequences for currents in RQM approaches are often ignored, 
beyond 4-momentum conservation which, of course, is always assumed. 
Quite generally, it supposes that currents transform as follows:
\begin{eqnarray}
e^{iP \cdot a}\;J^{\nu}(x) \;(S(x))\;e^{-iP \cdot a}=J^{\nu}(x+a) \;(S(x+a)),
\label{transl1}
\end{eqnarray}
where $P^{\mu}$ is the operator of the Poincar\'e algebra that generates 
space-time translations. In  the particular case $a=-x$, one gets:
\begin{eqnarray}
J^{\nu}(x) \;(S(x))=e^{iP \cdot x} \; J^{\nu}(0) \;(S(0))\;e^{-iP \cdot x}.
\label{transl2}
\end{eqnarray}
By considering the matrix element of the above current between eigenstates 
of momenta $P^{\mu}_i$ and $P^{\mu}_f$ and taking into account that 
these states in RQM approaches are, by construction 
\cite{Bakamjian:1953kh}, eigenstates of the operator  $P^{\mu}$, 
one can factorize the $x$ dependence of the current. When integrating 
the resulting factor, exp$(i\,(P_i\!-\!P_f) \!\cdot\! x)$, together with 
the plane wave describing the external probe, exp$(iq \!\cdot\! x)$, 
one recovers the energy-momentum conservation relation mentioned above, 
$P^{\mu}_f\!-\!P^{\mu}_i=q^{\mu} $. One is left with a matrix element at $x=0$.

\begin{figure}[htb]
\includegraphics[width=0.48\textwidth]{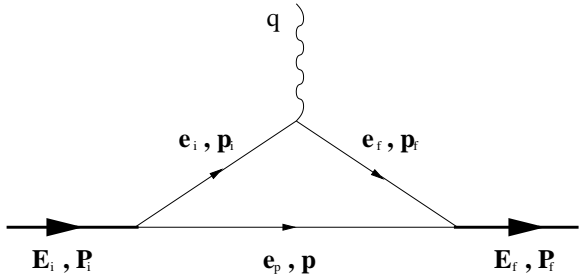}
 \vspace*{-8mm}
\caption{Interaction with an external field together with kinematical
notations.\label{fig2}}
\end{figure}

Quite generally however, writing relations given by Eqs. (\ref{transl1}) 
supposes that the currents in RQM approaches, beside a one-body component 
usually considered, also contain many-body components. In their absence,
relativistic covariance cannot be achieved. This can be checked 
by considering further relations stemming from Eqs. (\ref{transl1}) 
\cite{Lev:1993}:
\begin{eqnarray}
\Big[ P^{\mu}\;,\; J^{\nu}(x)\Big]=-i\partial^{\mu}\,J^{\nu}(x),
\;\;\;
\Big[ P^{\mu}\;,\; S(x)\Big]=-i\partial^{\mu}\,S(x), \label{comm1}
\end{eqnarray}
and especially the double commutator with $P^{\mu}$, which could 
be more relevant here: 
\begin{eqnarray}
\Big[P_{\mu}\;,\Big[ P^{\mu}\;,\; J^{\nu}(x)\Big]\Big]=
-\partial_{\mu}\,\partial^{\mu}\,J^{\nu}(x),
\;\;\; 
\Big[P_{\mu}\;,\Big[ P^{\mu}\;,\; S(x)\Big]\Big]=
-\partial_{\mu}\,\partial^{\mu}\,S(x)\, .\label{comm2}
\end{eqnarray}
Considering the matrix element of this last relation, it is found 
that the double commutator at the l.h.s. can be replaced by the square 
of the 4-momentum transferred to the system, $(P_i-P_f)^2=q^2$, 
while the derivatives at the r.h.s., for a single-particle current, 
can be replaced by the square of the 4-momentum transferred to the 
constituents, $(p_i-p_f)^2$ (see Fig. \ref{fig2} for both a graphical
representation and some kinematical notations). A test of Poincar\'e 
space-time translation covariance is therefore given by the relation at $x=0$: 
\begin{eqnarray}
<\;|q^2\; J^{\nu}(0) \;({\rm or}\;S(0))|\;>=
<\;|(p_i-p_f)^2\,J^{\nu}(0)\;({\rm or}\;S(0))|\;> \,.\label{comm3}
\end{eqnarray}
In RQM approaches, the squared momentum transferred to the
constituents differs most often from that one transferred to the system:
$(p_i-p_f)^2 \neq q^2\;(=-Q^2)$. It is therefore expected that 
Eq. (\ref{comm3}) is not fulfilled. Checking this relation, we found a close
relationship between its violation (a factor 3 for D.P.F., 30 for 
F.F.+I.F.(parallel) and 35000 for ``P.F." at the highest value of $Q^2$ 
considered here) and discrepancies between form factors. 

\begin{figure}[htb]
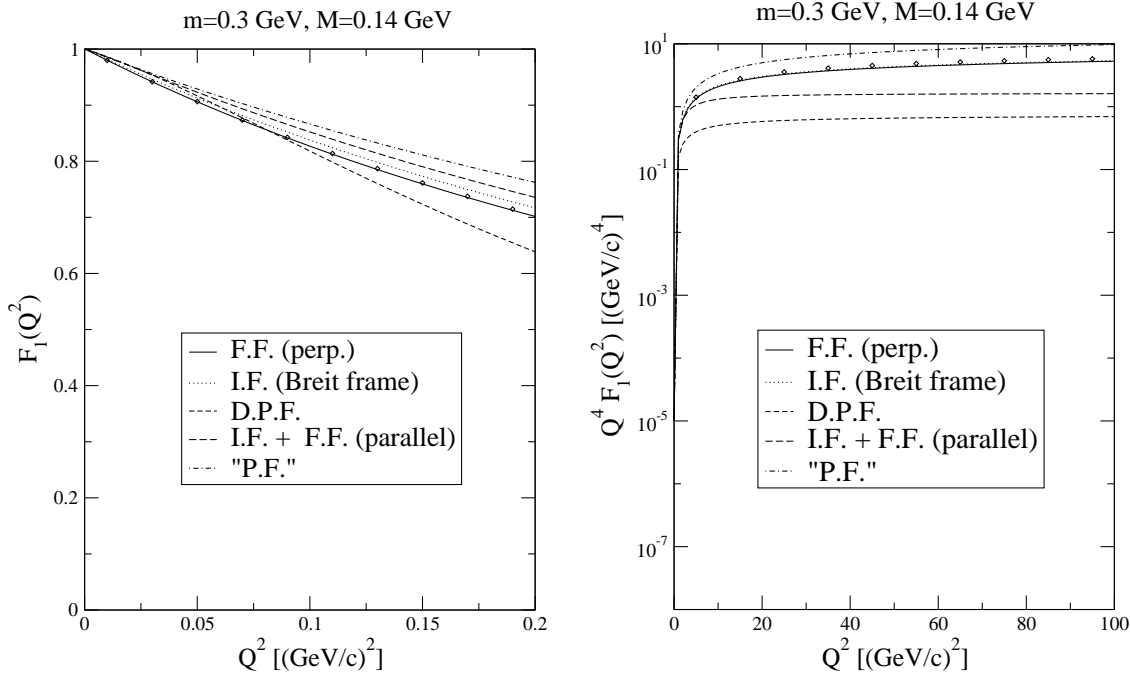

\includegraphics[width=0.45\textwidth]{DesplanquesB-fig2a.eps}
 \hspace*{4mm}\includegraphics[width=0.45\textwidth]{DesplanquesB-fig2b.eps}
\caption{Same as in Fig. \ref{fig1}, corrected for Poincar\'e space-time 
translation invariance.\label{fig3}}
\end{figure}

To correct for the violation in a first step,  we modified the coefficient 
of $Q$ in the boost expression so that to fulfill  the equality 
$(p_i-p_f)^2 =q^2.$
This can be done analytically in some cases, numerically (on the average) 
in other ones. The change amounts to account for many-body currents.
Interestingly, no modification is required for the standard front form 
($q^{+}=0$),  where the above equality is always fulfilled. 
Results so obtained are shown in Fig. \ref{fig3}. It is found that the
undesirable dependence of the charge radius on $M$  has vanished 
and that many orders of magnitude discrepancies at high $Q$ are largely removed.

Concerning a possible extension of present results, we recently found quite
similar ones for the charge pion form factor, indicating that the spin of the
constituents is not essential in the discussion made here \cite{Dong}. 
We also found 
that remaining discrepancies between different approaches (F.F.(perp.), 
I.F. (Breit frame), ``P.F." and a dispersion-relation one \cite{Krutov} 
corrected for some part) could be completely removed by relying 
on a current that differs from the usual choice. The modification has 
a small numerical effect but is important in getting the above result. 
This last achievement 
supposes a somewhat non-trivial change of variables relating one approach 
to the other, after accounting for effects from space-time translation 
invariance as described previously, when analytically possible. 
Both the numerical results and the recent developments unambiguously 
show the relevance of correctly implementing Poincar\'e space-time 
translation invariance.
\section{CONCLUSION AND OUTLOOK}
We have shown that large discrepancies between predicted form factors 
in different RQM approaches could be ascribed to missing properties 
from Poincar\'e space-time translation invariance. Relations that 
could be used to test this symmetry as well as to correct for its violation 
have been emphasized. Discrepancies are largely removed after this is done.
Some of them could be related to a scaling of the  charge radius with 
the inverse of the mass of the system under consideration. This 
counterintuitive result points to the violation of some symmetry. 
It is removed when accounting for Poincar\'e space-time translation invariance.
In comparison with Lorentz invariance, whose consequences can easily be checked, 
those related to the other symmetry discussed here are more subtle, 
perhaps explaining why they have not received much attention till now.

Present results point to the standard front-form or instant-form approaches 
as more efficient ones to implement relativity in describing properties 
of few-body systems. Va\-ri\-ous considerations were favoring these approaches 
but a deep and simple justification was lacking. We believe that fulfilling 
constraints from Poincar\'e space-time translation invariance is an 
essential argument in discriminating between different approaches. 

When comparing theoretical predictions to experiment, one could wonder 
about the reliability of either the approach used to implement relativity 
or the dynamical ingredients. By reducing the uncertainty about relativity, 
the present work allows one to concentrate the discussion on the role 
of the underlying dynamics.

\end{document}